\documentclass[aps,amsmath,twocolumn,amssymb,floatfix,showpacs,superscriptaddress,nofootinbib,longbibliography]{revtex4-1}
\usepackage{mathtools}
\usepackage{braket}
\usepackage[dvipsnames]{xcolor}
\usepackage{float}
\usepackage{subfigure}
\usepackage[colorlinks=true,linktoc=page,linkcolor=ForestGreen,citecolor=magenta,urlcolor=purple]{hyperref}
\newcommand{\vect}[1]{\boldsymbol{\mathrm{#1}}}
\mathchardef\mhyphen="2D % Define a "math hyphen"

\newcommand{\ie}{{i.e.,\,\,}}
\newcommand{\eg}{{e.g.,~}}
\newcommand{\ua}{{\uparrow }}
\newcommand{\da}{{\downarrow }}

\newcommand\bea{\begin{eqnarray}}
\newcommand\eea{\end{eqnarray}}
\newcommand\beq{\begin{equation}}  
\newcommand\eeq{\end{equation}}

\newcommand{\noi}{\noindent}

\newcommand{\non}{\nonumber}  
\newcommand{\dis}{\displaystyle}  
\usepackage[normalem]{ulem}
\definecolor{lime}{HTML}{A6CE39}
\usepackage{sidecap,tikz}
\DeclareRobustCommand{\orcidicon}{\hspace{-1.0mm}
	\begin{tikzpicture}
		\draw[lime, fill=lime] (0.0,0.0) 
		circle [radius=0.15] 
		node[white] {{\fontfamily{qag}\selectfont \tiny \,ID}};
		\draw[white, fill=white] (-0.0525,0.095) 
		circle [radius=0.007];
	\end{tikzpicture}
	\hspace{-3.0mm}
}
\foreach \x in {A, ..., Z}{\expandafter\xdef\csname orcid\x\endcsname{\noexpand\href{https://orcid.org/\csname orcidauthor\x\endcsname}
		{\noexpand\orcidicon}}
}

\AtBeginDocument{%
	\newwrite\bibnotes
	\def\bibnotesext{Notes.bib}
	\immediate\openout\bibnotes=\jobname\bibnotesext
	\immediate\write\bibnotes{@CONTROL{REVTEX41Control}}
	\immediate\write\bibnotes{@CONTROL{%
			apsrev41Control,author="08",editor="1",pages="1",title="1",year="1"}}
	\if@filesw
	\immediate\write\@auxout{\string\citation{apsrev41Control}}%
	\fi
}%
\begin{document}

%=============START of MAIN PAPER===============

\title{Engineering anomalous Floquet Majorana modes and their time evolution in helical Shiba chain}

\author{Debashish Mondal\orcidD{}}
\email{debashish.m@iopb.res.in}
\affiliation{Institute of Physics, Sachivalaya Marg, Bhubaneswar-751005, India}
\affiliation{Homi Bhabha National Institute, Training School Complex, Anushakti Nagar, Mumbai 400094, India}

\author{Arnob Kumar Ghosh\orcidA{}}
\email{arnob@iopb.res.in}
\affiliation{Institute of Physics, Sachivalaya Marg, Bhubaneswar-751005, India}
\affiliation{Homi Bhabha National Institute, Training School Complex, Anushakti Nagar, Mumbai 400094, India}

\author{Tanay Nag\orcidB{}}
\email{tanay.nag@physics.uu.se}
\affiliation{Department of Physics and Astronomy, Uppsala University, Box 516, 75120 Uppsala, Sweden}

\author{Arijit Saha\orcidC{}}
\email{arijit@iopb.res.in}
\affiliation{Institute of Physics, Sachivalaya Marg, Bhubaneswar-751005, India}
\affiliation{Homi Bhabha National Institute, Training School Complex, Anushakti Nagar, Mumbai 400094, India}

%--------------------------------------------------------
%--------------------------------------------------------
\begin{abstract}
\noi We theoretically explore the Floquet generation of Majorana end modes~(MEMs) (both regular $0$- and anomalous $\pi$-modes) implementing a periodic sinusoidal modulation in chemical potential in an experimentally feasible setup based on a one-dimensional chain of magnetic impurity atoms having spin spiral configuration (out-of-plane N\'eel-type) fabricated on the surface of most common bulk $s$-wave superconductor. We obtain a rich phase diagram in the parameter space, highlighting the possibility of generating multiple $0$-/$\pi$-MEMs localized at the end of the chain. We also study the real-time evolution of these emergent MEMs, especially when they start to appear in the time domain. These MEMs are topologically characterized by employing the dynamical winding number. We observe that the existing perturbative analysis is unable to explain the numerical findings, indicating the complex mechanism behind the formation of the Floquet Shiba minigap, which is characteristically distinct from other setup \eg Rashba nanowire model. We also discuss the possible experimental parameters in connection to our model. Our work paves the way to realize the Floquet MEMs in a magnet-superconductor heterostructure.
\end{abstract}
%--------------------------------------------------------
%--------------------------------------------------------

\maketitle

%======================================================
%\section{Introduction}
%======================================================
\textcolor{blue}{\textit{Introduction.}---} Majorana zero-modes~(MZMs) associated with topological superconductors~(TSCs)~\cite{Kitaev_2001,Kitaev2009,qi2011topological,Leijnse_2012,Alicea_2012,beenakker2013search,ramonaquado2017} have been attracting massive attention due to their non-Abelian braiding property, 
which is proposed to be the elementary building block for the fault tolerant topological quantum computations~\cite{Ivanov2001,freedman2003topological,KITAEV20032,Stern2010,NayakRMP2008}. 
The idea of MZMs was first proposed by Kitaev through the model of one-dimensional~(1D) spinless $p$-wave superconductor~\cite{Kitaev_2001,Kitaev2009}. However, this model is not experimentally feasible due to the unavailability of a 1D $p$-wave superconductor in nature. Nevertheless, there exists an alternate realistic proposal to engineer 1D Kitaev-like physics in 1D semiconducting nanowire (NW) with strong spin-orbit coupling~(SOC), placed in close proximity to a bulk $s$-wave superconductor in presence of a Zeeman field~\cite{Oreg2010,LutchynPRL2010,Leijnse_2012,Alicea_2012,Mourik2012Science,das2012zero,ramonaquado2017}. Majorana zero bias peak, observed in several transport experiments based on hybrid superconductor-semiconductor NW setups, has been interpreted as the indirect signatures of the MZMs~\cite{Mourik2012Science,das2012zero,Rokhinson2012,Finck2013,Albrecht2016,Deng2016}.

In recent times, the hunt for MZMs has taken an alternative route based on helical spin chain~\cite{Yazdani2013,Felix_analytics,Loss2013PRL,Felix2014,PhysRevB.89.115109, Bena2015,Loss2016,Christensen2016,Simon2017,Sticklet2019,Rex2020,Mohanta2021, Loss2022} or magnetic adatoms fabricated on the surface of a bulk $s$-wave superconductor
~\cite{Sau2013,PhysRevB.88.180503,Hui2015,Joel2015,Sharma2016,Theiler2019,Mashkoori2020,Teixeria2020}. Physically, the scattering between magnetic impurities and the quasiparticles in the superconductor fosters the formation of Yu-Shiba-Rusinov~(YSR) states~\cite{Shiba_original,Loss2021,Felix2021,Ortuzar2022,PritamPRB2023} inside the superconducting gap. These YSR states can hybridize among themselves and form YSR/Shiba bands. The helical spin texture and strength of the magnetic impurities play the combined role of SOC and magnetic field, respectively~\cite{Bena2015,Loss2022}. A few experimental proposals based on the scanning tunneling microscope (STM) measurement, have also been reported to realize the MZMs associated with the 
minigap of YSR bands~\cite{Yazdani_science,Yazdani2015,HowonSciAdv2018,Schneider2020,Wiesendanger2021,Beck2021,Wang2021PRL,Schneider2022,Crawford2022}.

%~~~~~~~~~~~~~~~~~~~~~~~~~~~~~~~~~~~~~~~~~~~~~~~~~~~~~~~~~
% Fig 1
%~~~~~~~~~~~~~~~~~~~~~~~~~~~~~~~~~~~~~~~~~~~~~~~~~~~~~~~~~
\begin{figure}[]
	\centering
	\subfigure{\includegraphics[width=0.37\textwidth]{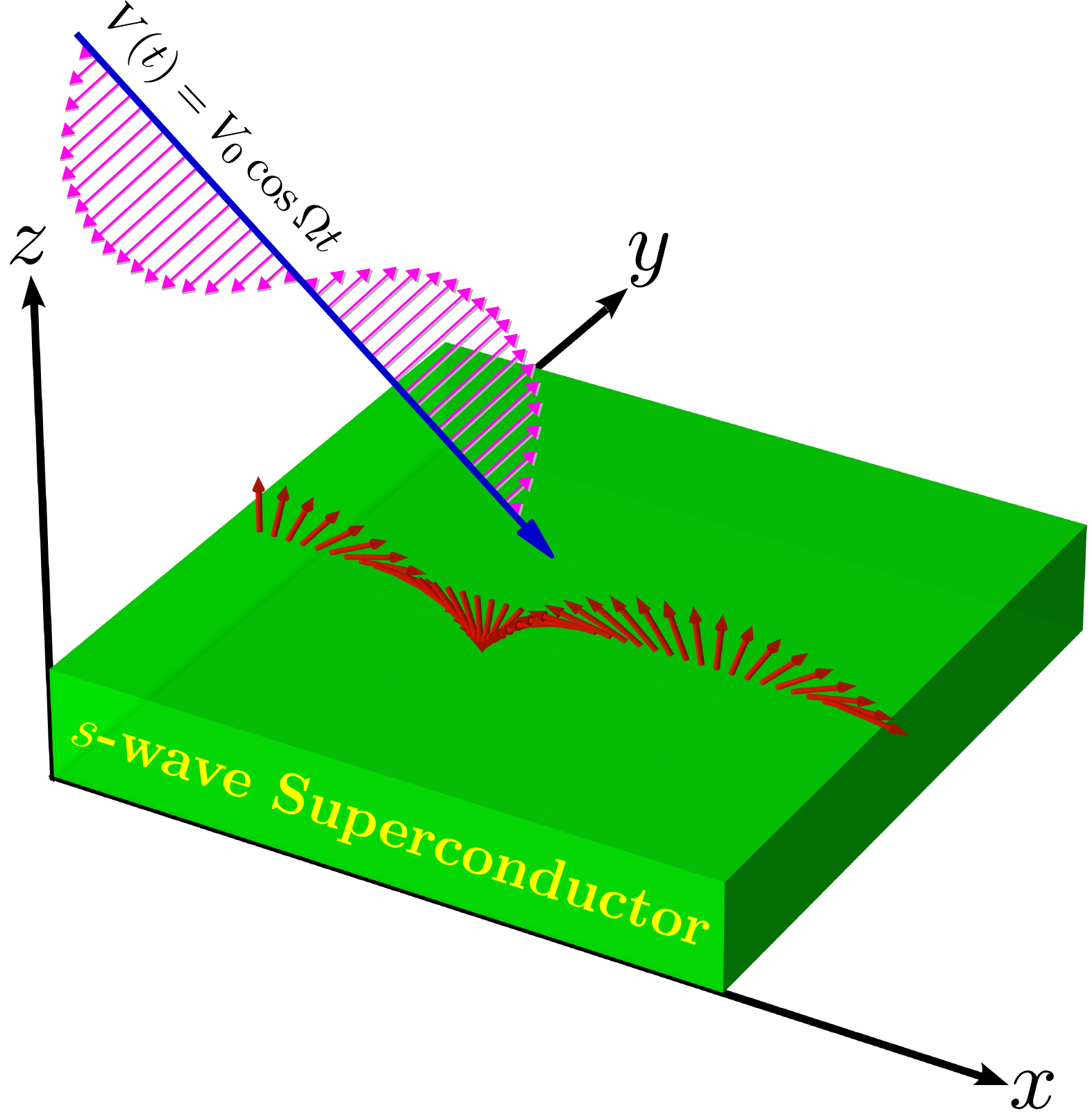}}
	\caption{Schematic representation of our setup is demonstrated here. A 1D chain of magnetic adatoms with their spins (red arrows) confined in the $xz$ plane (out-of-plane 
N\'eel-type spin spiral configuration) is placed on the top of a bulk $s$-wave superconductor (green).}
	\label{fig:geometry}
\end{figure} 
%~~~~~~~~~~~~~~~~~~~~~~~~~~~~~~~~~~~~~~~~~~~~~~~~~~~~~~~~~
%~~~~~~~~~~~~~~~~~~~~~~~~~~~~~~~~~~~~~~~~~~~~~~~~~~~~~~~~~

On the other hand, Floquet generation is a sophisticated and versatile way to engineer the on-demand topological phases out of a non-topological system~\cite{oka09photovoltaic,kitagawa11transport,lindner11floquet,Rudner2013,Usaj2014,Piskunow2014,Eckardt2017,Yan2017,oka2019,NHLindner2020,nag2021anomalous,ThakurathiPRB2013,benito14,PotterPRX2016,JiangPRL2011,ReynosoPRB2013,LiuPRL2013,ThakurathiPRB2017,YangPRL2021,Bao2022}. The absorption and emission of photons from the driving field lead to the formation of Floquet quasi-energy sidebands. These overlapping sidebands trigger the bandgap opening and band inversion resulting in the appearance of Floquet topological modes. The dynamical setup also facilitates the generation of the anomalous topological boundary modes at finite energy, namely $\pi$-modes, having no static counterpart. In this exciting direction of Floquet band engineering, the emergence of Floquet Majorana end modes (MEMs) has been explored in 1D $p$-wave Kitaev chain~\cite{ThakurathiPRB2013,benito14,PotterPRX2016}, 1D cold-atomic NW/$s$-wave superconductor heterostructure~\cite{JiangPRL2011,ThakurathiPRB2017,LiuPRL2013,YangPRL2021}, and also very recently, in realistic 1D Rashba NW Model~\cite{Mondal2023_NW}. The braiding of these Floquet boundary modes further adds merit to these systems due to their applicability in quantum computations~\cite{Bomantara18,BomantaraPRB2018,BelaBauerPRB2019,MatthiesPRL2022}. Notwithstanding, the Flquet generation of MEMs in realistic helical spin chain model along with their topological characterization is yet to be explored. At this stage, we would like to pursue the answers to the following intriguing questions: (a) Is it possible to generate and topologically characterize Floquet MEMs employing a realistic model based on a helical Shiba chain/$s$-wave superconductor (magnet-superconductor) heterostructure while starting from a trivial phase? (b) How do these dynamical MEMs within the emergent quasi-energy Shiba band evolve with time? 

\begin{figure}[]
	\centering
	\subfigure{\includegraphics[width=0.5\textwidth]{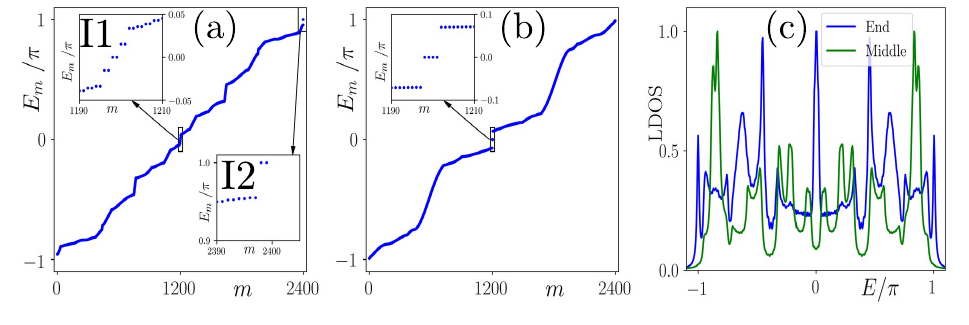}}
	\caption{We depict the quasi-energy spectra of the Floquet operator for $B/\Delta=2$ and $B/\Delta=3$ in panels (a) and (b), respectively. Floquet $0$- and $\pi$-MEMs are highlighted in the insets 
	I1 and I2. (c) We illustrate the energy resolved normalized LDOS computed at the end~(blue curve), and middle~(green curve) of the chain for $B/\Delta=2$. The green peaks represent the emergent Shiba modes within the Floquet quasi-energy spectrum. Here, we consider a 1D chain of 600 lattice sites and we choose the model parameters as ($\mu/\Delta, t_{h}/\Delta, \theta)=(4, 1, 2\pi/3)$ and $\Omega/\Delta=1.5$, $V_{0}/\Delta=5$.
	}
	\label{fig:dynamic}
\end{figure}

\begin{figure}[]
	\centering
	\subfigure{\includegraphics[width=0.47\textwidth]{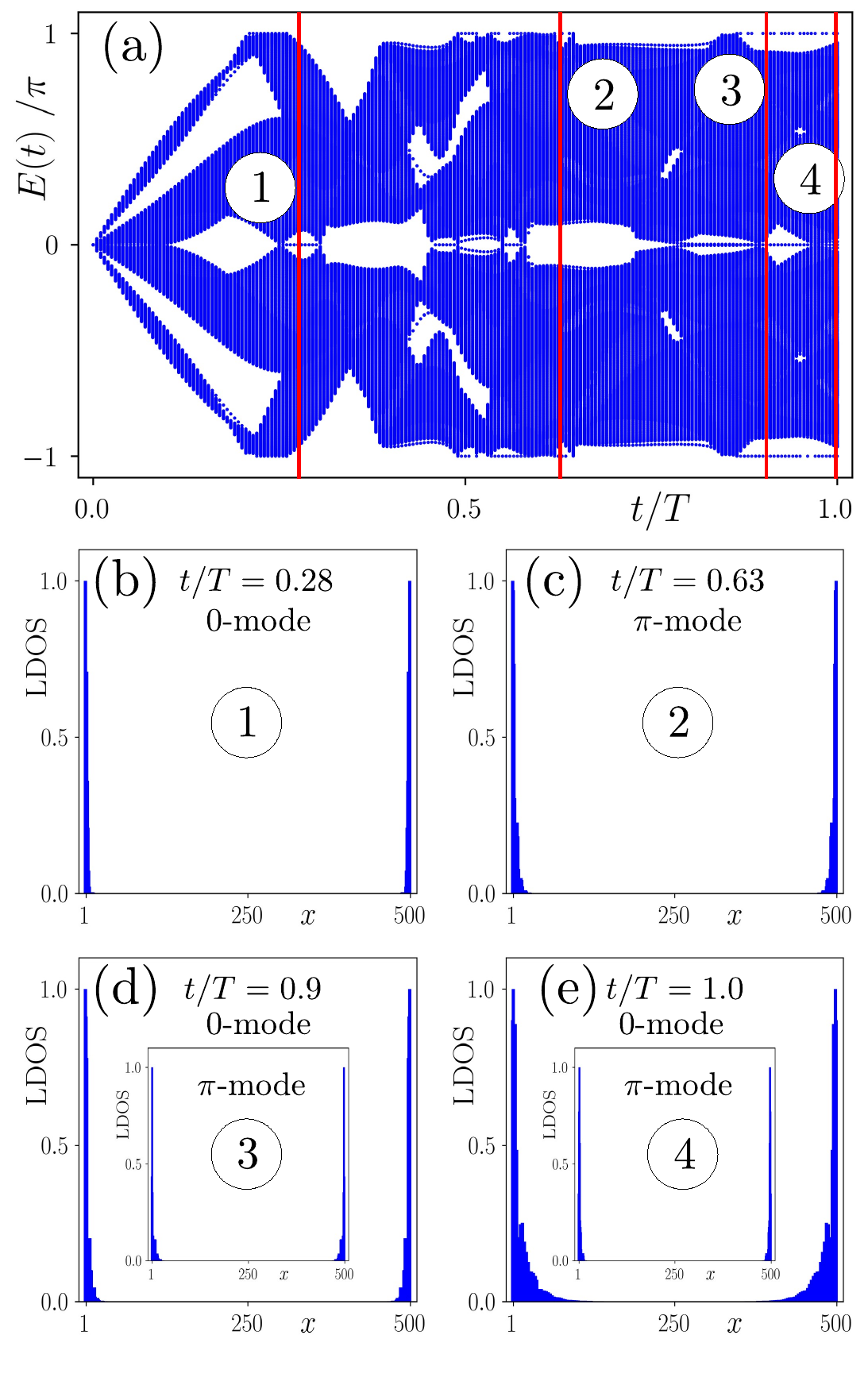}}
	\caption{In panel (a), we depict the time-dependent eigenvalue spectra of the time-evoltuion operator $U(t,0)$ as a function of time $t$. Both the $0$- and $\pi$-modes appear and disappear with time during the time-interval $t \in [0,T]$. We consider four time-points: $\textcircled{1}$ ($t=0.28T$) with only $0$-modes, $\textcircled{2}$ ($t=0.63T$) with only $\pi$-modes, $\textcircled{3}$ ($t=0.9T$) with both $0$- and $\pi$-modes, and $\textcircled{4}$ ($t=T$) with both $0$- and $\pi$-modes. The LDOS for $0$- and/or $\pi$-modes corresponding to $\textcircled{1} \mhyphen \textcircled{4}$ 
are illustrated in panels (b)-(e). Clearly, at $t/T=0.28$ ($t/T=0.63$) only 0-modes ($\pi$-modes) are present, while at $t/T=0.9, 1$ both the $0$- and $\pi$-MEMs appear. Time-dependent site resolved LDOS for these above mentioned MEMs corresponding to those time-instances are shown in the panels (b), (c), (d), (e) respectively. Here, we consider a 1D chain of 500 lattice sites and all the model parameters take the value as mentioned in Fig.~\ref{fig:dynamic}.}
	\label{fig:evolution}
\end{figure}

\begin{figure}[]
	\centering
	\subfigure{\includegraphics[width=0.47\textwidth]{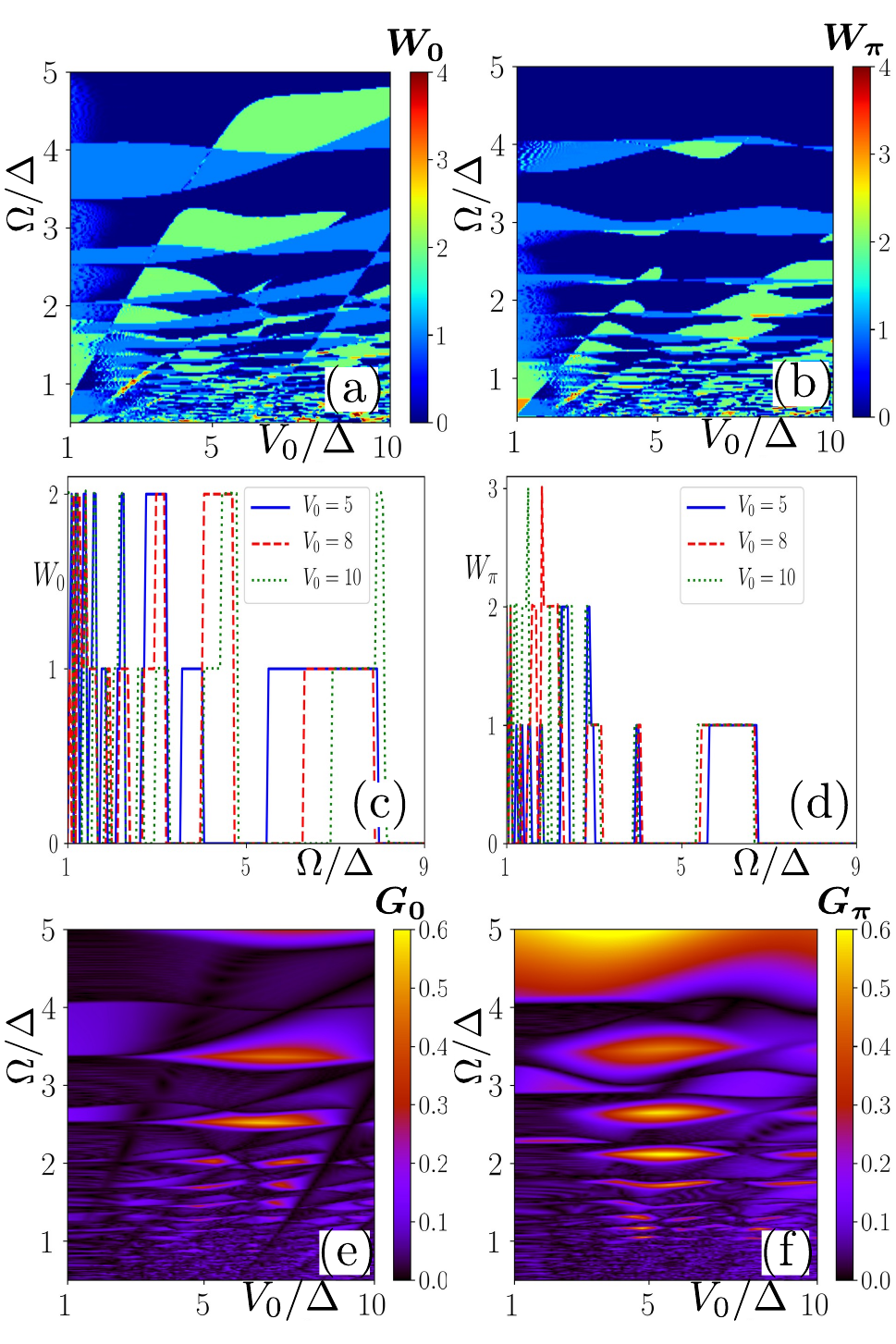}}
	\caption{We depict the dynamical winding number $W_{0}$ and $W_{\pi}$ in the $V_0 \mhyphen \Omega$ plane in the panels (a) and (b), respectively. Here, $W_{\epsilon}$ characterizes the Floquet MEMs residing at the quasi-enegy gap $\epsilon$. In the panels (c) and (d), we demonstrate $W_{0}$ and $W_{\pi}$ as a function of $\Omega$, respectively for three fixed values of $V_0/\Delta=$ 5~(solid blue line), 8~(dashed red line), and 10~(dashed green line) for better clarity. We illustrate the quasi-energy gap around quasi-energy $0$ and $\pi$ in panels (e) and (f), respectively. We choose $B/ \Delta=5$ (topological regime of the static model) and the rest of the model parameters take the same value as mentioned in Fig.~\ref{fig:dynamic}.}
	\label{fig:dynamical_winding}
\end{figure}

\begin{figure}[]
	\centering
	\subfigure{\includegraphics[width=0.48\textwidth]{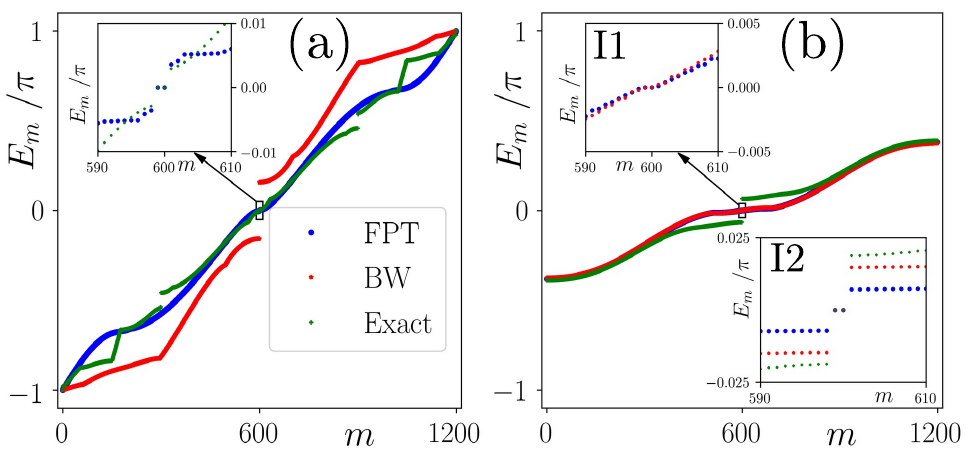}}
	\caption{We compare the eigenvalue spectra $E_m$ as a function of $m$, obtained from the FPT (blue dots), the BW perturbation theory (red dots), and the exact Floquet operator 
	(green dots). In panel (a), we depict the eigenvalue spectra for $(V_{0}/\Delta,\Omega/\Delta)=(10,3)$ while in the inset, we show the modes near quasi-energy zero. We choose $(V_{0}/\Delta,\Omega/
\Delta)=(5, 10)$ for panel (b) while the zoomed-in spectra near zero quasi-energy is shown in inset I1. In the inset I2, we demonstrate the eigenvalue spectra for $(V_{0}/\Delta,\Omega/\Delta)=(5, 6.5)$. 
	We choose 1D chain of 300 lattice sites and the rest of the model parameters same as mentioned in Fig.~\ref{fig:dynamic}. }
	\label{fig:analytics}
\end{figure}

In this article, we first briefly discuss about the underlying static model based on a helical 1D magnetic spin chain proximitized with a common bulk $s$-wave superconductor~\cite{Yazdani2013} (see Fig.~\ref{fig:geometry}). We explore the generation of Floquet MEMs (both $0$- and anomalous $\pi$-modes) employing an external periodic 
sinusoidal drive in this setup (see Fig.~\ref{fig:dynamic}). Afterward, we demonstrate the real-time evolution of the Floquet MEMs (see Fig.~\ref{fig:evolution}). We compute the dynamical winding number utilizing the periodized evolution operator to characterize the topological nature of the Floquet MEMs as shown in Fig.~\ref{fig:dynamical_winding}. We find that the numerically obtained quasi-energy spectrum cannot be described by the analytical perturbative analysis where the superconducting part is only renormalized without affecting the normal part in the Bogoliubov-de Gennes~(BdG) Hamiltonian (see Fig.~\ref{fig:analytics}). We also provide some probable experimental parameters concerning our system of interest.

%======================================================
%\section{Model}\label{Sec:II}
%======================================================
\textcolor{blue}{\textit{Static Model.}---} We consider the model of a helical spin-chain, based on a 1D chain of magnetic impurity atoms mimicking out-of-plane N\'eel-type spin spiral (SS) 
configuration~\cite{PritamPRB2023} that is fabricated on the surface of a bulk $s$-wave superconductor~\cite{Yazdani2013}~(see Fig.~\ref{fig:geometry} for the schematic). 
We employ the following BdG basis: 
$\Psi_{j}=\left\{c_{j \uparrow},c_{j \downarrow}, c_{j \downarrow}^{\dagger}, -c_{j \uparrow}^{\dagger} \right\}^{\textbf{t}}$; where, $c_{j \uparrow}^{\dagger}$~($c_{j \uparrow}$) and $c_{j \downarrow}^{\dagger}$~($c_{j \downarrow}$) represent quasiparticle creation (annihilation) operator for the spin-up and spin-down sector at a lattice site-$j$, respectively, and \textbf{t} indicates the transpose operation. Exploiting the BdG basis, we can write an effective 1D Hamiltonian in real space for our system as
\begin{widetext}
\begin{eqnarray}
H&=& \sum_{j} \Psi_{j}^\dagger \left[-\mu \Gamma_{1} +B \cos(j \theta) \Gamma_{2} +B \sin(j \theta) \Gamma_{3} + \Delta \Gamma_4 \right] \Psi_{j}+ \sum_{j}\Psi_{j,\eta}^\dagger t_{h}  \Gamma_{1} \Psi_{j+1}  +\  {\rm h.c.} 
\label{eq:static_1D_Hamiltonian} \ ,
\end{eqnarray}
\vspace{-0.2cm}
\end{widetext}
where, $\mu$, $B$, $\theta$, $\Delta$, and $t_{h}$ represent chemical potential, magnetic impurity strength, the angle between two adjacent spins, superconducting pairing gap, and the hopping amplitude, respectively. Also, $\Gamma_{1}=\tau_{z} \sigma_{0}$, $\Gamma_{2}=\tau_{0} \sigma_{z}$, $\Gamma_{3}=\tau_{0} \sigma_{x}$, $\Gamma_{4}=\tau_{x} \sigma_{0}$, while the Pauli matrices $\vect{\tau}$ and $\vect{\sigma}$ acts on the particle-hole and spin ($\uparrow$, $\downarrow$) subspaces, respectively. Here, we assume all the impurity spins as classical such that they are well separated from each other, and do not interact among themselves. The Hamiltonian $H$ [Eq.~(\ref{eq:static_1D_Hamiltonian})] preserves chiral symmetry $S= \mathbb{I}_{N} \otimes \tau_{y} \sigma_{y}$: $S^{-1} H S = - H$ and particle-hole symmetry~(PHS) $\mathcal{C}= \mathbb{I}_{N} \otimes \tau_{y} \sigma_{y} \mathcal{K}$: $\mathcal{C}^{-1}H\mathcal{C}=-H$; with $N$ and $\mathcal{K}$ representing the number of the impurity atoms and the complex-conjugation operation, respectively. However, $H$ breaks the time-reversal symmetry~(TRS) $\mathcal{T}= \mathbb{I}_{N} \otimes i \tau_{0} \sigma_{y} \mathcal{K}$: $\mathcal{T}^{-1} H \mathcal{T} \neq H$.

%~~~~~~~~~~~~~~~~~~~~~~~~~~~~~~~~~~~~~~~~~~~~~~~~~~~~~~~~~

%======================================================
%\section{Floquet generation}\label{Sec:III}
%======================================================
\textcolor{blue}{\textit{Floquet generation of MEMs.}---} We employ periodic sinusoidal drive in the on-site chemical potential to generate the Floquet TSC phase hosting MEMs. 
We choose the initial static Hamiltonian [Eq.~(\ref{eq:static_1D_Hamiltonian})] such that it resides in the trivial phase to begin with. The driving protocol reads
\begin{eqnarray}
V(t)= \sum_{j} \Psi_{j}^{\dagger}  \left[ V_{0} \cos (\Omega t) \Gamma_{1} \right] \Psi_{j}  \label{eq:drive_propocol}\ ,
\end{eqnarray}
where, $V_{0}$ and $\Omega(= 2\pi/T)$ represent the strengh and frequency of the drive, respectively; while $T$ stands for the time-period of the external drive. The periodicity of the drive $V(t+T)=V(t)$ is ensured in the full time-dependent Hamiltonian $H(t)=H+V(t)$ as $H(t+T)=H(t)$. We work in the real time-domain picture, and the time-evolution operator in terms of time-ordered~(TO) product reads as~\cite{AKG_time_evolution}
\begin{align}
U(t,0)&= \text{TO}  \exp \left[ - \int_{0}^{t} dt^{\prime} H(t^{\prime}) \right]  \ , \nonumber \\
&=\prod_{j=0}^{M-1} U(t_{j}+\delta t, t_{j}) \ , \label{eq:evolution_definiton}
\end{align}
where, $U(t_{j}+\delta t, t_{j})=e^{-i H(t_{j}) \delta t}$; with $\delta t = t/M$, $t_{j}=j \delta t$, and $M$ is the total number of Trotterization steps. The Floquet operator $U(T,0)$ is the time-evolution operator computed at $t=T$. Here, $U(T,0)$ serves the purpose of the dynamical analogue of the Hamiltonian. Thus, we diagonalize $U(T,0)$ to procure the quasi-energy spectrum for our system. The corresponding quasi-energy $E_m\in[-\pi,\pi]$. We demonstrate the quasi-energy spectrum as a function of the state index $m$ in Figs.~\ref{fig:dynamic} (a) and (b) for different impurity strengths ($B/\Delta=2$, and $B/\Delta=3$, respectively which indicate non-topological regime in the static model; see supplemental material~(SM)~\cite{supp}). For $B/\Delta=2$, we obtain two MEMs (one mode per end) at both quasi-energy $E=0$ and $\pi$ [see the insets I1 and I2 of Fig.~\ref{fig:dynamic}(a)]. While for $B/\Delta=3$, we highlight the generation of multiple MEMs at $E=0$ [see the inset of Fig.~\ref{fig:dynamic}(b)]. The generation of MEMs at quasi-energy $\pi$ is unique to the Floquet systems only, and does not have any static analog and this also serves as the prime result of the current manuscript considering Shiba chain model. Apart from $0$-energy, MEMs may also appear at finite energy ($\pm \pi$) in Floquet systems due to the presence of PHS in the underlying BdG Hamiltonian. Thus, the states having quasi-energy $0$ and $\pi \equiv -\pi$ can be their own antiparticle~\cite{JiangPRL2011}. In Fig.~\ref{fig:dynamic}(c), we depict the 
energy-resolved local density of states (LDOS) computed at the end (blue line), and middle (green line) of the chain. Note that, the LDOS is normalized by its maximum value throughout the manuscript. 
The peaks at $E=0,\pm \pi$, denoted by the blue curve, indicate the presence of Floquet MEMs when LDOS computed at the end of the chain. On the other hand, the green line in Fig.~\ref{fig:dynamic}(c), does not exhibit any peak at $E=0$ or $\pm \pi$ and it indicates the emergent bulk Floquet quasi-energy Shiba band within the superconducting gap as LDOS is calculated at the middle of the chain. 
Thus, the $0$- and $\pm \pi$-modes are truly boundary modes without having any weightage at the bulk. Moreover, the separation between the lowest green peaks (close to $0$- and 
$\pm \pi$) indicate the corresponding dynamical topological gap. 

%~~~~~~~~~~~~~~~~~~~~~~~~~~~~~~~~~~~~~~~~~~~~~~~~~~~~~~~~~
% Fig 3
%~~~~~~~~~~~~~~~~~~~~~~~~~~~~~~~~~~~~~~~~~~~~~~~~~~~~~~~~~
%~~~~~~~~~~~~~~~~~~~~~~~~~~~~~~~~~~~~~~~~~~~~~~~~~~~~~~~~~
%Time evolution \label{sec:III}
%~~~~~~~~~~~~~~~~~~~~~~~~~~~~~~~~~~~~~~~~~~~~~~~~~~~~~~~~~

\textcolor{blue}{\textit{Time evolution of the Floquet MEMs.}---} The real-time dependence of the driving protocol motivates us to pursue the time-evolution of the emergent Floquet MEMs~\cite{AKG_time_evolution}. To this end, we diagonalize $U(t,0)$, and illustrate the instantaneous eigenvalue spectra $E(t)$ as a function of time $t$ in Fig.~\ref{fig:evolution}(a). At $t=0$, the system
is a trivial superconductor. Afterwards, we observe that both the $0$- and $\pi$-MEMs continue to appear and disappear in the Floquet TSC phase as a function of time $t$ before one reaches the 
full time-period $T$ and the same can be identified distinctly from Fig.~\ref{fig:evolution}(a). To unveil the footprints of the MEMs at different time-instances, we consider four time-slices: $\textcircled{1}$ at $t=0.28T$ with only $0$-modes, $\textcircled{2}$ at $t=0.63T$ with only $\pi$-modes, $\textcircled{3}$ at $t=0.9T$ with both $0$- and $\pi$-modes, and $\textcircled{4}$ at $t=T$ when both $0$- and $\pi$-modes present. The LDOS associated with $0$- and/or $\pi$-modes corresponding to the points $\textcircled{1}\mhyphen\textcircled{4}$ are demonstrated with respect to the length of the chain in Figs.~\ref{fig:evolution}(b)-(e), respectively. It is worth mentioning here that both the $0$- and $\pi$-modes are sharply localized at the two ends of the 1D chain.

%~~~~~~~~~~~~~~~~~~~~~~~~~~~~~~~~~~~~~~~~~~~~~~~~~~~~~~~~
%Fig 4
%~~~~~~~~~~~~~~~~~~~~~~~~~~~~~~~~~~~~~~~~~~~~~~~~~~~~~~~~
%~~~~~~~~~~~~~~~~~~~~~~~~~~~~~~~~~~~~~~~~~~~~~~~~~~~~~~~~~
%~~~~~~~~~~~~~~~~~~~~~~~~~~~~~~~~~~~~~~~~~~~~~~~~~~~~~~~~~

%~~~~~~~~~~~~~~~~~~~~~~~~~~~~~~~~~~~~~~~~~~~~~~~~~~~~~~~~
%Fig 5
%~~~~~~~~~~~~~~~~~~~~~~~~~~~~~~~~~~~~~~~~~~~~~~~~~~~~~~~~
%~~~~~~~~~~~~~~~~~~~~~~~~~~~~~~~~~~~~~~~~~~~~~~~~~~~~~~~~~
%~~~~~~~~~~~~~~~~~~~~~~~~~~~~~~~~~~~~~~~~~~~~~~~~~~~~~~~~~

%======================================================
%\section{Topological characterization}\label{Sec:IV}
%======================================================
\textcolor{blue}{\textit{Topological characterization of the dynamical MEMs.}--} The broken translational symmetry in $H$ [Eq.~(\ref{eq:static_1D_Hamiltonian})] rules out the computation of topological invariant in momentum space. However, one can employ twisted boundary condition~(TBC) in real space geometry to topologically characterize the system~\cite{Mondal2023_NW}. To this end, we connect the two ends of the 1D chain and embed a hypothetical flux $\eta$ through it. The flux induces a Peierls phase substitution to the hopping amplitude as $t_h \rightarrow t_h e^{i \eta j}$ with $j \in \mathbb{Z}$~\cite{Mondal2023_NW}. The Hamiltonian $H$ [Eq.~(\ref{eq:static_1D_Hamiltonian})] as well as the time-evolution operator $U(t,0)$ [Eq.~(\ref{eq:evolution_definiton})] become an explicit function of $\eta$ such that: $H \rightarrow H(\eta)$ and $U(t,0) \rightarrow U(\eta;t,0)$. We enforce periodic boundary condition~(PBC) employing the constraint on $\eta$ as $N \eta=2 \pi$. Here, we note that depending upon the angle between two successive impurity spin $\theta$, the PBC can be achieved only for a few specific values of $N$ (see SM~\cite{supp} for details). To characterize the MEMs with a distinct topological invariant, we also need to invoke the periodized evolution operator, constructed employing TBC, defined as~\cite{Rudner2020,PhysRevB.104.L140502,GhoshPRB2022,supp}
\begin{eqnarray}
U_{\epsilon}(\eta;t,0)=U(\eta;t,0) [U(\eta;T,0)]_{\epsilon}^{-t/T} \label{eq:U_epsilon}.
\end{eqnarray}
Here, $\epsilon$ stands for the $0$- and $\pi$-gap. Afterward, we exploit the chiral symmetry operator to define a dynamical winding number $W_\epsilon$ to topologically characterize the MEMs appearing at quasi-energy $\epsilon$~\cite{Mondal2023_NW,supp}, as
\begin{equation}
W_{\epsilon}= \left| \frac{i}{2 \pi} \int_{-\pi}^{\pi} \! \! d\eta \ {\rm Tr} \left[\left\{U_{\epsilon}^\pm(\eta;T/2,0)\right\}^{-1} \partial_{\eta}  U_{\epsilon}^\pm(\eta;T/2,0)\right] \right| \! .
\label{eq:dynamical_wind} 
\end{equation}
One can obtain $U_{0}^\pm(\eta;T/2,0)~[U_{\pi}^\pm(\eta;T/2,0)]$ by writing the periodized evolution operator employing chiral basis (see SM~\cite{supp} for details). Here, $W_{\epsilon}$ counts 
the number of Floquet MEMs in the TSC phase residing at a particular quasi-energy $\epsilon$. 

We depict the dynamical winding number $W_{0}$ and $W_{\pi}$ in the $V_0 \mhyphen \Omega$ plane in Figs.~\ref{fig:dynamical_winding}(a) and (b), respectively; while the color bar represents the number of Floquet MEMs. In certain parameter spaces, one can notice that the number of $0$- and/or $\pi$-mode becomes more than one. For better clarity, we also illustrate $W_{0}$ and $W_{\pi}$ as a function of $\Omega$ for three different fixed values of $V_{0}$ in Figs.~\ref{fig:dynamical_winding}(c), and (d), respectively. Thus, the appearance of multiple Floquet MEMs is also supported by the outcome of the dynamical winding number. We further show the bulk gap at quasi-energy $0$ ($G_0$) and $\pi$ ($G_\pi$) in the $V_0 \mhyphen \Omega$ plane in Figs.~\ref{fig:dynamical_winding}(e) 
and (f), respectively. Each gap closing in $G_0$ and $G_\pi$ correspond to a jump in $W_{0}$ and $W_{\pi}$ respectively, indicating a topological phase transition. This one-to-one mapping between the dynamical winding number $W_\epsilon$ and the bulk gap $G_\epsilon$ is evident from Fig.~\ref{fig:dynamical_winding}.

%======================================================
%\section{Perturbation theory}\label{Sec:IVB}
%======================================================

%~~~~~~~~~~~~~~~~~~~~~~~~~~~~~~~~~~~~~~~~~~~~~~~~~~~~~~~~
%Fig 6
%~~~~~~~~~~~~~~~~~~~~~~~~~~~~~~~~~~~~~~~~~~~~~~~~~~~~~~~~
%~~~~~~~~~~~~~~~~~~~~~~~~~~~~~~~~~~~~~~~~~~~~~~~~~~~~~~~~~

\textcolor{blue}{\textit{Perturbative analysis.}---} Having investigated the problem numerically, we employ two perturbative schemes such as the Floquet perturbation theory~(FPT)~\cite{Bhaskar_Mukherjee_2020,Sen_2021,S_Ghosh_2022,AKG_time_evolution} and the Brillouin-Wigner~(BW) perturbation theory~\cite{Mikami_BW_2016,GhoshPRB2020}, to examine the high-amplitude driving $V_0 \gg t_{h}$, and the high-frequency driving $\Omega \gg t_h$, respectively. We find that the superconducting term is renormalized in the leading order as follows $\Delta_{\rm FPT}^{\rm eff}=  \Delta \ J_{0}\left( \frac{2 V_{0}}{\Omega}\right)$ and $\Delta_{\rm BW}^{\rm eff}=  \Delta \ \left( 1- \frac{V_{0}^2}{2 \Omega^2}\right)$ for FPT and BW techniques, respectively; where $J_{0}(x)$ is the $0^{\rm {th}}$ Bessel function of the first kind (see SM~\cite{supp} for details). Notably, the normal part of the BdG Hamiltonian remains unaffected by such perturbations in the leading order.

We compute the quasi-energy spectrum in the real space for the effective Floquet Hamiltonian. In Fig.~\ref{fig:analytics}, we compare the quasi-energy spectra obtained from the FPT Hamiltonian (blue dots), BW Hamiltonian (red dots), and the exact Floquet operator (green dots). We demonstrate the cases where FPT/BW can successfully predict the zero-energy modes for very few choices of parameters. However, FPT/BW fails to anticipate the zero-energy mode and the overall quasi-energy profile for most of the parameter space, even though the above theories are applicable to those parameter choices. In Fig.~\ref{fig:analytics} (a), we depict an instant where FPT matches with the exact numerics to a certain extent; interestingly, the zero-energy modes are successfully predicted. 
Although, the BW severely fails to mimic the exact spectrum [see the inset of Fig.~\ref{fig:analytics} (a) for better clarity]. On the other hand, in Fig.~\ref{fig:analytics} (b), FPT and BW exhibit similar spectrum while the exact numerics is drastically different as FPT/BW (exact numerics) exhibits gapless (gapped) feature around zero quasi-energy [see I1 of Fig.~\ref{fig:analytics} (b) for better clarity]. However, as we lower the drive amplitude and increase the drive frequency, FBT/BW can sometimes mimic the zero energy modes as noticed from the exact numerics [see I2 of Fig.~\ref{fig:analytics} (b)]. The match between theory and numerics is not rigorous but only accidental. Hence, neither the FPT nor BW perturbation theory can satisfactorily explain the emergence of MEMs in the driven system.

The numerical findings apprehend that the drive and magnetic impurity both simultaneously modify normal and superconducting parts of the BdG Hamiltonian. Since the above perturbation theories do not modify the normal part of the BdG Hamiltonian, indicating that the FPT/BW cannot capture the modulation of the Shiba minigap in the presence of the time-periodic drive. Note that, the emergent Floquet Shiba minigap comprises of both the renormalized superconducting and the normal parts. More precisely, the closed form expression of the dynamical Shiba gap is not tractable as the driving term, namely, chemical potential, is hidden inside the static Shiba bands in a complex manner. Finding an appropriate theory to understand the Floquet Shiba minigap in a driven helical spin chain 
is far more complex than anticipated and beyond the scope of the current manuscript. Interestingly, our numerical investigation with the quasi-energy spectrum indicates that the driven chain starts hosting Floquet MEMs for substantially lower strengths of magnetic impurity as compared to the static case~\cite{supp}.

%======================================================
%\section{Plausible experimental realization}\label{Sec:V}
%======================================================
\textcolor{blue}{\textit{Plausible experimental realization.}---} Having demonstrated the generation of Floquet MEMs theoretically, here we discuss the possible way to realize our setup experimentally. 
The most suitable superconducting candidate material can be the bulk Nb (110) since it has the highest possible superconducting gap of $\Delta=1.51$ meV among the conventional superconductors~\cite{Wiesendanger2021}. Afterward, one may fabricate magnetic adatoms such as Mn/Cr over the bulk Nb (110) employing STM-based single-atom manipulation methods~\cite{Eigler1990,HowonSciAdv2018,Schneider2020,Schneider2022}. This method could provide better tunability of the angle between two impurity spins. Following our model, we consider the scenario described in Fig.~\ref{fig:dynamic}(a), and to this end, the other model parameters can take the values: $t_h=\Delta=1.51$ meV, $B=2\Delta=3.02$ meV, and $\mu=4 \Delta=6.04$ meV. Moreover, the periodic sinusoidal modulation of the on-site chemical potential can be achieved through an AC gate voltage having an amplitude of $V_0=5\Delta=7.55$ meV and frequency of $\Omega \approx 3.44$ THz for the generation of Floquet $0$- and $\pi$ MEMs.

%======================================================
%\section{Summary and conclusion}\label{Sec:VI}
%======================================================
\textcolor{blue}{\textit{Summary and conclusion.}---} To summarize, in this article, we demonstrate an experimentally feasible way to engineer Floquet MEMs employing a helical spin-chain model 
(in the form of N\'eel-type SS configuration fabricated on the surface of an $s$-wave superconductor) and periodic modulation of chemical potential. We obtain a rich phase diagram in the parameter 
space, allowing us to realize both $0$- and $\pi$-Floquet MEMs in the emergent quasi-energy band. We also investigate the time evolution and tunability of the number of Floquet MEMs. The $0$- and 
$\pi$-MEMs are always sharply localized at the two ends of the system. We topologically characterize the MEMs utilizing the dynamical winding number in a real-space picture. Our study indicates the limitation of perturbative calculations that only refers to the renormalization of the superconducting part in the BdG Hamiltonian in mimicking the exact numerical result. Therefore, our work reveals an angle where magnetic impurity physics gets intertwined with the Floquet drive to render the $p$-wave superconducting gap, which is extremely non-trivial to anticipate a priori. This complex phenomenon can be easily distinguished from the Floquet Rashba NW physics, where the normal part of the BdG Hamiltonian is only modified~\cite{Mondal2023_NW}. Our numerical findings clearly show the substantial reduction of the strength of the magnetic impurity to host the Floquet MEMs compared to that for the static case~\cite{supp}. However, we leave a detailed microscopic understanding of the emergent dynamical Shiba minigap ($0$ and $\pi$-gap) for future studies. We also provide probable experimental parameters to realize our setup. These Floquet MEMs are relatively robust as we start from the trivial phase and possibly can be probed by the STM signal.

%======================================================
%\subsection*{Acknowledgments}
%======================================================
\textcolor{blue}{\textit{Acknowledgements}---}D.M., A.K.G., and A.S. acknowledge SAMKHYA: High-Performance Computing Facility provided by Institute of Physics, Bhubaneswar, for numerical computations. D.M. and A.K.G. thank Pritam Chatterjee for stimulating discussions.

%\newpage

\bibliography{bibfile}{}

%============End of MAIN PAPER=============
%==============SUPPLEMENTARY=============
\normalsize\clearpage
\begin{onecolumngrid}
\begin{center}
	{\fontsize{12}{12}\selectfont
		\textbf{Supplementary Material for ``Engineering anomalous Floquet Majorana modes and their time evolution in helical Shiba chain''\\[5mm]}}
	{\normalsize Debashish Mondal\orcidD{},$^{1,2}$ Arnob Kumar Ghosh\orcidA{},$^{1,2}$ Tanay Nag\orcidB{},$^{3}$ and Arijit Saha\orcidC{},$^{1,2}$,\\[1mm]}
	{\small $^1$\textit{Institute of Physics, Sachivalaya Marg, Bhubaneswar-751005, India}\\[0.5mm]}
	{\small $^2$\textit{Homi Bhabha National Institute, Training School Complex, Anushakti Nagar, Mumbai 400094, India}\\[0.5mm]}
	{\small $^3$\textit{Department of Physics and Astronomy, Uppsala University, Box 516, 75120 Uppsala, Sweden}\\[0.5mm]}
	{}
\end{center}

\newcounter{defcounter}
\setcounter{defcounter}{0}
\setcounter{equation}{0}
\renewcommand{\theequation}{S\arabic{equation}}
\setcounter{figure}{0}
\renewcommand{\thefigure}{S\arabic{figure}}
\setcounter{page}{1}
\pagenumbering{roman}

\renewcommand{\thesection}{S\arabic{section}}
%\tableofcontents
%\textcolor{red}{
%~~~~~~~~~~~~~~~~~~~~~~~~~~~~~~~~~~~~~~~~~~~~~~~~~~~~~~~~	
\section{Topological phase boundaries of the static Hamiltonian and static winding number}
\label{Sec:Static_winding_number}
%~~~~~~~~~~~~~~~~~~~~~~~~~~~~~~~~~~~~~~~~~~~~~~~~~~~~~~~~
%~~~~~~~~~~~~~~~~~~~~~~~~~~~~~~~~~~~~~~~~~~~~~~~~~~~~~~~~~
% Fig 2
%~~~~~~~~~~~~~~~~~~~~~~~~~~~~~~~~~~~~~~~~~~~~~~~~~~~~~~~~~
\begin{figure}[H]
	\centering
	\subfigure{\includegraphics[width=0.97\textwidth]{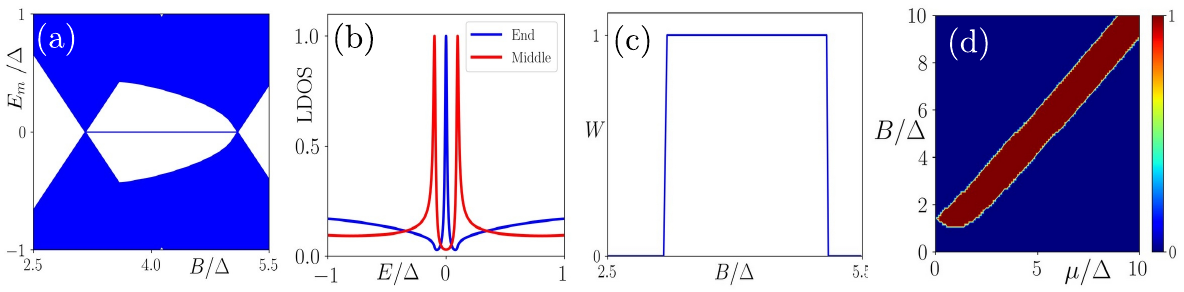}}
	\caption{(a) We depict the energy eigenvalue spectra for the Hamiltonian [see Eq.~(1) in the main text or Eq.~(\ref{eq:sup_static_hamiltonian})] employing OBC as a function of the magnetic impurity strength $B/\Delta$. (b) We illustrate the energy resolved LDOS spectra computed at the end (blue curve), and middle (red curve) of the magnetic chain in the topoplogical regime ($B/\Delta=5$). (c) We depict the winding number $W$ corresponding to (a). We obtain $W=1$ for $B_{-} \leq B \leq B_{+}$, and zero otherwise. (d) We illustrate winding number $W$ in the $\mu$-$B$  plane. The red~(blue) region indicates the topological (non-topological) regime with $W=1~(0)$ [Eq.~(\ref{eq:sup_static_hamiltonian})]. We choose the model parameters as ($\mu/\Delta, t_{h}/\Delta, \theta)=(4, 1, 2\pi/3)$.
	%are taken to be the same as in Fig.~2 in the main text.
	}
	\label{fig:sup_static_winding_number}
\end{figure} 
%~~~~~~~~~~~~~~~~~~~~~~~~~~~~~~~~~~~~~~~~~~~~~~~~~~~~~~~~~
%~~~~~~~~~~~~~~~~~~~~~~~~~~~~~~~~~~~~~~~~~~~~~~~~~~~~~~~~~
The static Hamiltonian, that we introduce in Eq.~(1) in the main text, can also be recast in the following form
\begin{align}
	H=N_{1} \Gamma_{1} + N_{2} \Gamma_{2} + N_{3} \Gamma_{3} + N_{4} \Gamma_{4}  \  ,
	\label{eq:sup_static_hamiltonian}
\end{align}
with $\dis N_{1}=-\mu \sum_{i=1}^{N} \psi_{i}^{\dagger} \psi_{i} +t_{h} \sum_{i=1}^{N-1} \psi_{i}^{\dagger} \psi_{i+1} + h.c. $ , $\dis N_{2}=B \sum_{i=1}^{N} \cos (i \theta) \psi_{i}^{\dagger} \psi_{i}$, $\dis N_{3}=B \sum_{i=1}^{N} \sin (i \theta) \psi_{i}^{\dagger} \psi_{i}$, $\dis N_{4}=\Delta \sum_{i=1}^{N} \psi_{i}^{\dagger} \psi_{i}$, and $N$ being the number of lattice sites and $\Gamma_{1}, \Gamma_{2}, \Gamma_{3}, \Gamma_{4}$ have been already defined in the main text. Considering the impurity strength $B$ as a parameter, we investigate the topological phase boundaries of the static Hamiltonian $H$. Owing to the 1D nature of the system, one can compute the Pfaffian for the static system analytically (see Sec.~\ref{Sec:Pffaffian} and also Ref.~\cite{Yazdani2013}) and trace out the range of $B$ for which the system is topological as: $B_-< \lvert B \rvert < B_+$; with $B_{\pm}=\sqrt{\Delta^{2}+(|\mu|\pm 2 \cos(\theta/2)t_{h})^{2}}$.
%The system resides in the topological regime in the parameter space when the Pfaffian is negative~\cite{Yazdani2013,supp}. Employing the negativity condition of the Pfaffian, one can trace out the range %of $B$ for which the system is topological as: $B_-< \lvert B \rvert < B_+$; with $B_{\pm}=\sqrt{\Delta^{2}+(|\mu|\pm 2 \cos(\theta/2)t_{h})^{2}}$. 
Note that, $B_{\pm}$ reduces to the usual expression as obtained for the NW set up when $\theta=0$~\cite{Oreg2010}. To verify the appearance of the MZMs, we employ open boundary condition~(OBC) and illustrate the energy eigenvalue spectra of $H$ [Eq.~(\ref{eq:sup_static_hamiltonian})] as a function of $B$ in Fig.~\ref{fig:sup_static_winding_number}(a). We find that the MZMs appear when 
$B_-< \lvert B \rvert < B_+$. 
%In the topological regime, we depict the energy-resolved local density of states~(LDOS) computed at the middle and end of the chain in Fig.~\ref{fig:sup_static_winding_number}(b). 
%\textcolor{red}{Note that the LDOS is normalized by its maximum value throughout the manuscript.} 
In the topological regime, when the  local density of states~(LDOS) is calculated at the end (middle) of the chain, we obtain the signatures of the MZMs~[represented by the blue curve in 
Fig.~\ref{fig:sup_static_winding_number}(b)] (the YSR/Shiba peaks only [represented by the red curve in Fig.~\ref{fig:sup_static_winding_number}(b)]). %The peaks appearing in the red curve are referred to as the Shiba peaks, 
The energy separation between the two lowest Shiba peaks designates the mini-gap.

To establish the topological nature of the MZMs, we also compute the winding number utilizing the chiral symmetry operator.  In the real space, to compute the winding number, we employ twisted boundary condition~(TBC)~\cite{Demler2010PRB,Titum2016PRX,Sreejith2016,Mondal2023_NW} by connecting two ends of the magnetic chain together to form a ring with a hypothetical peridic flux $\eta$ being threaded through the ring. In this hypothetical setup, a particle acquires a phase whenever it hops from one site to another site~\cite{Niu1984,NiuPRB1985}. This refers to the following transformation~\cite{Mondal2023_NW}: $\psi_{j,  \ua (\da)} \rightarrow e^{i \eta j }\psi_{j,\eta \ua (\da)}$ and $\psi_{j,  \ua (\da)}^\dagger \rightarrow e^{-i \eta j }\psi_{j,\eta \ua (\da)}^\dagger$ such that $N \eta=2 \pi$. In TBC, the Hamiltoian $H$ [Eq.~(\ref{eq:sup_static_hamiltonian})] takes the following form
\begin{eqnarray}
H_{\rm twist}(\eta)&=& \sum_{j,\eta} \Psi_{j,\eta}^\dagger \big[-\mu \Gamma_{1} +B \cos(j \theta) \Gamma_{2} +B \sin(j \theta) \Gamma_{3} + \Delta \Gamma_4 \big]  \Psi_{j,\eta}+ \sum_{j,\eta}\Psi_{j,\eta}^\dagger t_{h} e^{i \eta} \Gamma_{1} \Psi_{j+1,\eta}  +\  {\rm h.c.} \label{eq:sup_twisted_hamiltonian}
\end{eqnarray}
The chiral symmetry operator $S$ anti-commutes with the Hamiltoian $H_{\rm twist}(\eta)$ [Eq.~(\ref{eq:sup_twisted_hamiltonian})] \ie $\{S,H_{\rm twist }(\eta)\}=0$. This indicates that chiral basis 
$U_{S}$ which diagonalizes $S$ would anti-diagonalize $H_{\rm twist}(\eta)$. Thus, $H_{\rm twist }(\eta)$ in the chiral basis reads as
\begin{eqnarray}
\tilde{H}=U^{\dagger}_{S} H_{\rm twist}(\eta) U_{S}= \begin{pmatrix}
0 &\tilde{H}^{+}(\eta)\\
\tilde{H}^{-}(\eta)&0
\end{pmatrix} \ , \label{eq:sup_H_tilde}
\end{eqnarray}
where, $\tilde{H}^{\pm}(\eta)$ denote $2N\times 2N$ squre matrices that are defined on $\pm$ chiral block, respectively. We can define the chiral winding number $W$ employing 
$\tilde{H}^{\pm}(\eta)$ as~\cite{RyuNJP2010,ChiuRMP2016,Mondal2023_NW}
\begin{equation}
W=\Bigg| \pm \frac{i}{2 \pi} \int_{- \pi}^{\pi} d\eta \hspace{0.1 cm} {\rm Tr} \left[\left\{\tilde{ H}_{\rm twist}^{\pm}(\eta)\right\}^{-1} \partial_{\eta} \tilde{ H}_{\rm twist}^{\pm}(\eta)\right] \Bigg|
\label{eq:sup_static_wind}.
\end{equation}
We compute this chiral winding number $W$ by employing the above equation and depict it's behavior as a function of $B$ in Fig.~\ref{fig:sup_static_winding_number} (c). A non-zero $W$ within the topological phase boundary $B_{-}\leq B \leq B_{+}$ exhibits one to one correspondence with the energy eigenvelue specta shown in Fig.~\ref{fig:sup_static_winding_number} (a). We also illustrate $W$ in the $\mu$-$B$ plane in Fig.~\ref{fig:sup_static_winding_number}(b) to highlight the topological superconducting (TSC) region within the static Shiba chain model.

	%~~~~~~~~~~~~~~~~~~~~~~~~~~~~~~~~~~~~~~~~~~~~~~~~~~~~~~~~
	\section{Analytical approach to calculate the topological phase boundary}
	\label{Sec:Pffaffian}
	The topological phase boundary for the static Hamiltonian [Eq.~(1) in the main text or Eq.~(\ref{eq:sup_static_hamiltonian})] can also be computed analytically employing the Pfaffian invariant. The detailed procedure to compute the Pfaffian invariant has been discussed in Ref.~\cite{Yazdani2013}. Here, we mention only a few steps to compute the same. The static lattice Hamiltonian for the magnetic impurity chain consisting of magnetic atoms placed on the top of a bulk $s$-wave superconductor can also be written as
	\begin{eqnarray}
	H_{\rm lattice}=&&\sum_{n,\alpha} t_{h} f_{n,\alpha}^{\dagger} f_{n+1,\alpha} -\sum_{n,\alpha} \mu f_{n,\alpha}^{\dagger} f_{n,\alpha} + \sum_{n,\alpha,\beta} (\vec{B}_{n}.\vec{\sigma})_{\alpha \beta} f_{n,\alpha}^{\dagger} f_{n,\beta} + \sum_{n} \Delta f_{n,\uparrow}^{\dagger} f_{n,\downarrow}^{\dagger} + {\rm h.c.}\ ,
   \label{eq:sup_lattice_Hamiltonian}
	\end{eqnarray}
	where, $n$ is the site-index, and $\alpha$ and $\beta$ are the spin-indices. Here, all the magnetic atoms have same moment but different orientation: $\vec{B}_{n}=B(\sin \theta_{n} \cos \phi_{n} \hat{i}+\sin \theta_{n} \sin \phi_{n} \hat{j}+\cos \theta_{n} \hat{k})$. Using an appropriate unitary transformation one can make the magnetic moments diagnonal in $\uparrow \downarrow$ basis~\cite{Beenakker2011,Yazdani2013}. Without loss of generality, we restict the spin orientations in the $xz$ plane in the form of out-of-plane N\'eel-type spin spiral configuration~\cite{PritamPRB2023}). 
	One can write down the first quantized Hamiltonian in the momentum space, employing a Majorana basis as~\cite{Yazdani2013}
	\begin{eqnarray}
		H(p)=
		\begin{pmatrix}
			0 & 0 & 2 t_{h} \cos \left(\frac{\theta}{2}\right) \cos p + B - \mu &  2 i t_{h} \sin \left(\frac{\theta}{2}\right) \sin p - \Delta\\
			0 & 0 & -2 i t_{h} \sin \left(\frac{\theta}{2}\right) \sin p + \Delta & 2 t_{h} \cos \left(\frac{\theta}{2}\right) \cos p - B - \mu\\
			-2 t_{h} \cos \left(\frac{\theta}{2}\right) \cos p - B + \mu & 2 i t_{h} \sin \left(\frac{\theta}{2}\right) \sin p - \Delta & 0 & 0\\
			-2 i t_{h} \sin \left(\frac{\theta}{2}\right) \sin p + \Delta &   -2 t_{h} \cos \left(\frac{\theta}{2}\right) \cos p + B + \mu  & 0  &  0	
		\end{pmatrix} \ . \nonumber\\ 
		\label{eq:sup_Momemtum_Hamiltonian_elements}
	\end{eqnarray}
	The above Eq.~(\ref{eq:sup_Momemtum_Hamiltonian_elements}) indicates that $\theta$ is $4 \pi$ periodic, while $p$ is $2 \pi$ periodic, and the periodicity condition for the Hamiltonian 
	can be achieved when both $\theta$ and $p$ are periodic, or both of them are anti-periodic due to the presence of functions like $\sin (\theta/2) \sin p$ and $\cos (\theta/2) \cos p$ 
	in Eq.~(\ref{eq:sup_Momemtum_Hamiltonian_elements}). Let $\theta=\frac{2 \pi}{N}N_{h}$, with $N$ ($N_{h}$) being the total number of sites (rotations by helix from $n=1$ 
	to $n=N$). Depending upon different values of $N$ and $N_{h}$, one can acheive the periodicity condition in four ways (for $j \in \mathbb{N}$)~\cite{Yazdani2013}:
	\begin{eqnarray}
	&& \text{1. $p=\frac{2 \pi}{N}j$  \hspace*{2 mm} if mod$[N_{h},N]\leq \frac{N}{2}$ and $N_{h}$ even.} \nonumber \\
	&& \text{2. $p=\frac{ \pi}{N}(2j+1)$  \hspace*{2 mm} if mod$[N_{h},N]\leq \frac{N}{2}$ and $N_{h}$ odd.} \nonumber\\
	&& \text{3. $p=\frac{2 \pi}{N}j$  \hspace*{2 mm} if mod$[N_{h},N]> \frac{N}{2}$ and $[N-\text{mod}[N_{h},N],2]=0$.} \nonumber \\
	&& \text{4. $p=\frac{ \pi}{N}(2j+1)$  \hspace*{2 mm} if mod$[N_{h},N]> \frac{N}{2}$ and $[N-\text{mod}[N_{h},N],2]=1$} \label{eq:sup_p_values}.
	\end{eqnarray}
	Note that, $p=0,~\pi$ are allowed only when both $N$ and $N_{h}$ are even. For example, for $\theta=\frac{2 \pi}{3}$, one needs to satisfy $\frac{N}{N_{h}}=3$. In this scenario, to obtain the 
	periodic boundary condition, we need to consider $N_h=2j$ and $N=6j$; with $j \in \mathbb{N}$.
	
	Now the Pfaffian for the skew-symmetric matrix $H(p)$ reads as
	\begin{eqnarray}
	Pf[H(0)]=B^{2}-[\Delta^{2}+(\mu+2 \cos(\theta/2)t_{h})^2], \hspace{2 mm} Pf[H(\pi)]=B^{2}-[\Delta^{2}+(\mu+2 \cos(\theta/2)t_{h})^2] \label{eq:sup_Pfaffian}\ .
	\end{eqnarray}
	The system becomes topologically non-trivial, when $\left\{{\rm sign}(Pf[H(0)]) \times {\rm sign} (Pf[H(\pi)])\right\}$ is negative~\cite{Yazdani2013}. Thus, we find that the system represents a 
	topological superconducting phase when the following condition is satisfied:
	\begin{eqnarray}
	\sqrt{\Delta^{2}+(|\mu|-2 |\cos(\theta/2)t_{h}|)^2} <|B|< \sqrt{\Delta^{2}+(|\mu|+2 |\cos(\theta/2)t_{h}|)^2}\ .
	\label{eq:sup_topological_criteria}
	\end{eqnarray}
	This [Eq.~(\ref{eq:sup_topological_criteria})] gives rise to the topological phase boundaries $B_{\pm}$ as mentioned in the main text.

	%~~~~~~~~~~~~~~~~~~~~~~~~~~~~~~~~~~~~~~~~~~~~~~~~~~~~~~~~
	\section{Dynamical winding number}
	\label{Sec:dynamical_winding_number}
	For the dynamical case, to calculate topological index, time evolution operator carries out the same job done by Hamiltonian for the static case. This happens only for the case when any one type of dymanical modes (either $0$- or $\pi$-modes) is present. If both type of modes appear in the quasi-energy spectrum, then normal time evolution operator only calculates the difference between the number of modes present at quasi-energies $0$ and $\pi$~\cite{Rudner2020}. To count the actual number of modes at each energy level, we need to introduce the of notion of gap $\epsilon$ via the periodized evolution operator as~\cite{Rudner2013,GhoshPRB2022,Mondal2023_NW}
	\begin{eqnarray}
	U_{\epsilon}(\eta,t)=U(\eta,t) [U(\eta,T)]_{\epsilon}^{-t/T} \label{eq:sup_U_epsilon}\ ,
	\end{eqnarray}
	where, $U(\eta,t)$ stands for the time-evolution operator of the Hamiltonian, $H_{\rm twist}(\eta)$ [Eq.~(\ref{eq:sup_twisted_hamiltonian})] and $[U(\eta,T)]_{\epsilon}^{-t/T}$ is the $(-t/T)^{\rm th}$ power of Floquet operator for the $\epsilon$- gap which is given by
	\begin{eqnarray}
	\left[U_{\rm step}(\eta,T)\right]_{\epsilon}^{-t/T}&=&  \sum_{n=1}^{L/2} e^{-i(2 \pi -2\epsilon-|E_{n}(\eta)|)t/T} \ket{\psi_n(\eta)} \bra{\psi_n(\eta)} + \sum_{n=L/2 +1}^{L} e^{- i\left|E_{n}(\eta) \right|t/T} \ket{\psi_n(\eta)} \bra{\psi_n(\eta)}
	\label{eq:sup_uth}, \quad 
	\end{eqnarray}
	where, $n=1 \cdots L/2$ ($L/2+1 \cdots N$) stands for valence~(conduction) bands. Note that, $U_{\epsilon}(\eta,0)=U_{\epsilon}(\eta,T)=\mathbb{I}$ and $U_{\epsilon}(\eta,t)$ is periodic in time 
	\ie $U_{\epsilon}(\eta,t)=U_{\epsilon}(\eta,t+T)$. Moreover, chiral symmetry of the system imposes the following constraint on the periodized 
	evolution operator~\cite{Michel2016,PhysRevB.105.064304} as
	\begin{equation}\label{eq:sup_constraintsPEO}
	S^{-1} U_{\epsilon}(\eta,t) S= U_{-\epsilon}(\eta,-t) e^{2 \pi i t /T} \ . 
	\end{equation} 
	At the half period ($t=\frac{T}{2}$), periodicity condition of $U_{\epsilon}(k,t)$ reads the above constraint as: $S U_{\epsilon}\left(\eta,\frac{T}{2}\right) S^{-1}=-U_{-\epsilon}\left(\eta,\frac{T}{2}\right)$.   
	Hence, for the $0$- and $\pi$-gaps, that becomes
	\begin{eqnarray}
	S^{-1} U_{0}\left(\eta,\frac{T}{2}\right) S=-U_{0}\left(\eta,\frac{T}{2}\right)  \  , \hspace*{2 mm}
	S^{-1} U_{\pi}\left(\eta,\frac{T}{2}\right) S=U_{\pi}\left(\eta,\frac{T}{2}\right)\ ,  \label{eq:sub_0picostraint}
	\end{eqnarray} 
	where, we use the relation $U_{-\pi}\left(\eta,\frac{T}{2}\right)=-U_{\pi}\left(\eta,\frac{T}{2}\right)$.  Thus, in the chiral basis, $U_{0}\left(\eta,\frac{T}{2}\right)$~$\left[U_{\pi}\left(\eta,\frac{T}{2}\right)
	\right]$ takes block anti-diagonal [diagonal] form as
	\begin{eqnarray}
	\tilde{U}_{0}\left(\eta,\frac{T}{2}\right)&=&U_S^{\dagger} \hspace*{1mm} U_{0}\left(\eta,\frac{T}{2}\right) U_S = 
	\begin{pmatrix}
	0& U_{0}^{+}(\eta)\\
	U_{0}^{-}(\eta)&0
	\end{pmatrix} , \  \ \quad
	\label{eq:sup_antidiag_U_zero} \\
	\tilde{U}_{\pi}\left(\eta,\frac{T}{2}\right)&=&U_S^{\dagger} \hspace*{1mm} U_{\pi}\left(\eta,\frac{T}{2}\right) U_S = 
	\begin{pmatrix}
	U_{\pi}^{+}(\eta)&0\\
	0&U_{\pi}^{-}(\eta)
	\end{pmatrix}
	\label{eq:sup_diag_U_pi}.
	\end{eqnarray}
	Therefore, using $U_{\epsilon}^{\pm}(\eta)$, the dynamical winding number $W_{\epsilon}$ for $\epsilon(=0,\pi)$-gap can be defined as~\cite{Demler2010PRB,Titum2016PRX,Sreejith2016}
	\begin{equation}
	W_{\epsilon}= \left| \pm \frac{i}{2 \pi} \int_{-\pi}^{\pi} d\eta \ {\rm Tr} \left[\left\{U_{\epsilon}^{\pm}(\eta)\right\}^{-1} \partial_{\eta}  U_{\epsilon}^{\pm}(\eta)\right] \right|
	\label{eq:sup_dynamical_wind}.
	\end{equation}
	Here, $W_{\epsilon}$ manifests itself as dynamical winding number by counting total number of modes present at one end of the chain at the quasi energy $\epsilon$.
%\textcolor{red}{
%~~~~~~~~~~~~~~~~~~~~~~~~~~~~~~~~~~~~~~~~~~~~~~~~~~~~~~~~
\section{Perturbative Analysis} \label{Sec:perturbation}
%~~~~~~~~~~~~~~~~~~~~~~~~~~~~~~~~~~~~~~~~~~~~~~~~~~~~~~~~
Here, we present the details of the Floquet perturbation theory~(FPT) and Brillouin-Wigner~(BW) perturbation theory results discussed in the main text for our set-up.
%------------------------------------------
\subsection{Floquet perturbation theory}
%------------------------------------------
The FPT provides us analytical understanding of a periodically driven system when the amplitude of the time-depedent part of the Hamiltonian is much larger than the that of the time-independent part~\cite{Bhaskar_Mukherjee_2020,Sen_2021,S_Ghosh_2022,AKG_time_evolution}. In the FPT formalism, the time-independent part is handled exactly and the time-dependent part is treated as perturbation. To this end, we write the full time-periodic Hamiltonian in the presence of harmonic drive as
\begin{equation}
	H(t)=N_{1} \Gamma_{1} + N_{2} \Gamma_{2}+N_{3} \Gamma_{3}  +N_{4} \Gamma_{4} + V_{0} \cos(\Omega t) \Gamma_{1}
	\label{eq:sub_inst_sup}\ ,
\end{equation}
where $\Gamma_{1}=\tau_{z} \sigma_{0}$, $\Gamma_{2}=\tau_{0} \sigma_{z}$, $\Gamma_{3}=\tau_{0} \sigma_{x}$, $\Gamma_{4}=\tau_{x} \sigma_{0}$. These $\vect{\Gamma}$ matrices satisfy the following relations: $[\Gamma_{1},\Gamma_{2}]=[\Gamma_{1},\Gamma_{3}]=[\Gamma_{2},\Gamma_{4}]=[\Gamma_{3},\Gamma_{4}]=0$  and $\{\Gamma_{1},\Gamma_{4}\}=\{\Gamma_{2},\Gamma_{3}\}=0$. We consider $V_{0} \gg t_h$ \ie amplitude of the drive is much larger than the hopping amplitude. Then, time-evolution operator for $V(t)$ reads as
\begin{equation}
	U_{0}(t,0)={\rm TO} \exp\left[-i \int_{0}^{t} dt^{\prime}V(t^{\prime})\right] = \exp\left[ -\frac{i V_{0}}{\Omega} \sin(\Omega t) \Gamma_{1} \right] 
	\label{eq:sup_u0_small_t}.
\end{equation}
One can notice that $U_{0}(T,0)=I$. We employ interaction picture and the time evolution operator for $H(t)$ is given as~\cite{Bhaskar_Mukherjee_2020,Sen_2021,S_Ghosh_2022,AKG_time_evolution}
\begin{equation}
	U_{I}(t,0)= I + (-i) \int_{0}^{t} dt^{\prime} H_{I}(t^{\prime})+ (-i)^{2} \int_{0}^{t}dt_{1} H_{I}(t_{1}) \int_{0}^{t_1}dt_{2} H_{I}(t_{2}) + .... \ ,
	\label{eq:sup_dyson} 
\end{equation}
where, $H_{I}(t)$ reads as
\begin{equation}
	H_{I}(t)=U_{0}^{\dagger}(t,0)  H U_{0}(t,0)
	\label{eq:sup_HI}.
\end{equation}
Following the relation bewteen the $\vect{\Gamma}$ matrices and the driving protocol, we obtain the following relations
\begin{align}
	e^{i A \Gamma_{1}} \Gamma_{4} e^{-i A \Gamma_{1}}&= e^{i 2 A \Gamma_{1}} \Gamma_{4} \non \\
	e^{i A \Gamma_{j}} \Gamma_{1} e^{-i A \Gamma_{j}} &= \Gamma_{j} \ , \qquad \qquad {\rm with} \ j=1,2,3.	 \label{gammarelation}
\end{align}
where, $A=\frac{i V_{0}}{\Omega} \sin(\Omega t)$. Using the Eqs.~(\ref{gammarelation}), we recast Eq.~(\ref{eq:sup_HI}) as 
\begin{equation}
	H_{I}(t)= \left( N_{1}\Gamma_{1} +N_{2}\Gamma_{2}+N_{3} \Gamma_{3}\right) +N_{4}\Gamma_{4} \exp\left[ \frac{i2 V_{0}}{\Omega} \sin(\Omega t) \Gamma_{1} \right] \ . \label{HamIrecast}
\end{equation}
Afterward we use the Jacobi-Anger Identity as
\begin{equation}
	e^{i z \sin x} =\sum_{n=-\infty}^{\infty} J_{n}(z) e^{i nz}=J_{0}(z)+\sum_{\{n\}\neq 0} J_{n}(z) e^{i nz} \label{eq:sup_Jacobi_Anger} \ ,
\end{equation}
and rewrite Eq.~(\ref{HamIrecast}) as
\begin{equation}
	H_{I}(t)= \left( N_{1}\Gamma_{1} +N_{2}\Gamma_{2}+N_{3} \Gamma_{3} +J_{0}\left(\frac{2 V_{0}}{\Omega}\right) N_{4}\Gamma_{4} \right) + N_{4}\Gamma_{4} \sum_{\{n\}\neq 0} J_{n}\left(\frac{2 V_{0}}{\Omega}\right) e^{in \Omega t} \label{eq:sup_HI_expression}.
\end{equation}
Now, following the FPT, the full time-evolution operator reads as~\cite{Bhaskar_Mukherjee_2020,Sen_2021,S_Ghosh_2022,AKG_time_evolution}
\begin{equation}
	U_{P}(t,0)=U_{0}(t,0)U_{I}(t,0) \ .	 \label{FPTto}
\end{equation}
Therefore, one can obtain the Floquet operator by replacing $t \rightarrow T$ in Eq.~(\ref{FPTto}) as
\begin{equation}
	U_{P}(T,0)=U_{0}(T,0)U_{I}(T,0) \ .	 \label{FPTFO}
\end{equation}
Thus, the Floquet operator $U_{P}(T,0)$ can be obtained using Eqs.~(\ref{eq:sup_dyson}), (\ref{eq:sup_HI_expression}), and (\ref{FPTFO}) as
\begin{align}
U_{P}(T,0)&=I + (-i) \int_{0}^{T} dt^{\prime} \hspace*{1 mm} H_{I}(t^{\prime}) + (-i)^{2} \int_{0}^{T} dt_{1} \hspace*{1 mm} H_{I}(t_{1}) \int_{0}^{t_{1}} dt_{2} \hspace*{1 mm} H_{I}(t_{2})+.... \non \\
&=I + U_{P}^{(1)}(T) +U_{P}^{(2)}(T)  + .... \label{eq:sup_U_seies} \ .
\end{align}
We only consider the leading order term in the perturbation series and obtain $U_{P}^{(1)}(T)$ as
\begin{equation}
U_{P}^{(1)}(T)=(-i)\left[   N_{1}\Gamma_{1} +N_{2}\Gamma_{2}+N_{3} \Gamma_{3} +J_{0}\left(\frac{2 V_{0}}{\Omega}\right) N_{4}\Gamma_{4}    \right]T \ ,
\end{equation}
Thus, the effective Hamiltonian, upto the leading order, reads as
\begin{equation}
H_{\rm FPT}^{\rm eff, (1)} = N_{1} \Gamma_{1} + N_{2} \Gamma_{2}+N_{3} \Gamma_{3} + J_{0}\left(\frac{2V_{0}}{\Omega}\right) N_{4} \Gamma_{4} \label{eq:sup_leading_eff}\ .
\end{equation}
Hence, employing the FPT and considering only the leading order term, we obtain only a 
modification to the bare superconducting gap as
\begin{equation}
\Delta_{\rm FPT}^{\rm eff}=\Delta \ J_{0}\left(\frac{2V_{0}}{\Omega} \right) \ .
\label{eq:sup_Delta_eff_FPT}
\end{equation}
%--------------------------------------------------
\subsection{Brillouin-Wigner perturbation theory}
%---------------------------------------------------
In the high-frequency limit \ie the drive frequency $\Omega \gg t_h ({\rm{hopping~amplitude}})$, 
we can obtain a closed form Hamiltonian employing perturbation theory such as Brillouin-Wigner~(BW) theory~\cite{Mikami_BW_2016,GhoshPRB2020}. Using BW perturbation theory, one can obtain the BW effective Hamiltonian as ~\cite{Mikami_BW_2016,GhoshPRB2020}
\begin{equation}
	H_{\rm BW}^{\rm eff}=\sum_{n=0}^{\infty} H_{\rm BW}^{(n)} \ ,
\end{equation}
with
\begin{align}
	H_{\rm BW}^{(0)}&=H_{0,0} \ ,  \non  \\
	H_{\rm BW}^{(1)}&= \sum_{\{n_1\}\neq 0} \frac{H_{0,n_1}H_{n_1,0}}{n_1 \Omega} \ , \non \\
	H_{\rm BW}^{(2)}&= \sum_{\{n_i\}\neq 0} \frac{H_{0,n_1} H_{n_1,n_2} H_{n_2,0}}{n_1 n_2 \Omega^2}-\frac{H_{0,n_1} H_{n_1,0} H_{0,0}}{n_{1}^2\Omega^2}  \ , \non \\
	H_{\rm BW}^{(3)}&=\mathcal{O} \left(\frac{1}{\Omega^3}\right) \ .
\end{align}
Here, the Fourier components $H_{m,n}$ read as
\begin{equation}
	H_{m,n}=\frac{1}{T} \int_{0}^{T} dt H(t) e^{i (m-n)\Omega t} \label{Fourier_Hamiltonian} \ .
\end{equation}
Following our driving protocol [Eq. (2) in the main text], we obtain
\begin{equation}
	H_{0,n}=\frac{V_{0}}{2}\Gamma_{1} \delta_{n,\pm 1}  \ . \label{Fcomp}
\end{equation}
Employing Eqs.~(\ref{Fourier_Hamiltonian}) and (\ref{Fcomp}), we obtain $H^{(0)}_{\rm BW}$, $H^{(1)}_{\rm BW}$, and $H^{(2)}_{\rm BW}$ as
\begin{align}
	H_{BW}^{(0)} &=H_{0,0}=H \non \ , \\
	H_{BW}^{(1)} &=\frac{V_{0}^2}{4 \Omega} I \non \ , \\
	H_{BW}^{(2)} &= - \frac{V_{0}^{2}}{2 \Omega^{2}} N_{4} \Gamma_{4} \ . \label{BW_terms}
\end{align}
Here, $H_{BW}^{(1)}$ only produces a trivial shift to the energy spectrum. Thus, we ignore this term and the BW Hamiltonian upto the leading non-vanishing order can be written as
\begin{equation}
	H_{\rm BW}^{\rm eff}= N_{1} \Gamma_{1} +N_{2} \Gamma_{2}+N_{3} \Gamma_{3} +N_{4}\left(1- \frac{V_{0}^2}{2 \Omega^{2}}\right) \Gamma_{4} \label{effective_Hamiltonian_BW} \ .
\end{equation}
Thus, similar to the FPT, the BW theory also provides a modification to the parent superconducting 
term as
\begin{equation}
	\Delta_{\rm BW}^{\rm eff}=\Delta  \left(1- \frac{V_{0}^2}{2 \Omega^{2}}\right)  
	\label{Delta_eff_BW}.
\end{equation}
On the other hand, if we consider the following limit: $\frac{V_{0}}{\Omega} \ll 1$, then we can obtain 
$J_{0}\left(\frac{2 V_{0}}{\Omega}\right) \approx \left(1 - \frac{V_{0}^2}{\Omega^2}\right)$. Hence, 
$\Delta_{{\rm{FPT}}}^{{\rm{eff}}} \approx \Delta_{{\rm{FPT}}}^{{\rm{BW}}}$ in such regime.

%~~~~~~~~~~~~~~~~~~~~~~~~~~~~~~~~~~~~~~~~~~~~~~~~~~~~~~~~
\section{MEMs in quasi-1D case: Static and Dynamic} \label{Sec:quasi1D}
%~~~~~~~~~~~~~~~~~~~~~~~~~~~~~~~~~~~~~~~~~~~~~~~~~~~~~~~~

%-------------------------	
\subsection{Static case}
%-------------------------
In the main text, we have restricted our discussion to a strictly 1D setup. However, in the realistic scenario, the parent superconductors are always with dimension more than one. Thus, to imitate the practical situation, we consider a quasi-1D system \ie 1D magnetic chain deposited on the top of a two-dimensional (2D) $s$-wave superconductor with kinetic energy~\cite{Yazdani2013} 
[see Fig.~1(a) in the main text]. To be precise, we stack the 1D chains in the orthogonal direction with the same matrix elements connecting the adjacent chains as considered for the longitudinal direction. The impurity is only considered in the middle chain of this stacking pattern. We consider the Hamiltonian of this quasi-1D system accordingly and choose all the parameters same as strictly 1D case and diagonalize the Hamiltonian. To identify the topological phase boundary, we depict its energy eigenvalue spectra as a function of $B$ in Fig~\ref{fig:sup_quasi_1D}(a). Interestingly, we observe a slight shift of the topological phase boundary in the parameter space compared to the strictly 1D case [see Fig.~2(a) in the main text and Fig.~\ref{fig:sup_quasi_1D}(a)]. However, the qualitaive nature of the Shiba band remains same as before. We also compute the site-resolved local density of states~(LDOS) for zero-energy MEMs in the topological [non-topological] regime and illustrate it in Figs.~\ref{fig:sup_quasi_1D}(b), [(c)]. In the topological regime, the zero-energy Majorana modes are sharply localized at the end of the impurity chain. In the non-topological regime, however, we do not observe any LDOS associated with zero energy at the end of the chain as there is no state present at $E_m=0$.
Here, the finite LDOS appears due to the bulk Shiba states inside the superconducting gap. Hence, the quasi-1D system exhibits qualitatively similar results as the strictly 1D case.
%~~~~~~~~~~~~~~~~~~~~~~~~~~~~~~~~~~~~~~~~~~~~~~~~~~~~~~~
\begin{figure}[]
	\centering
	\subfigure{\includegraphics[width=0.9\textwidth]{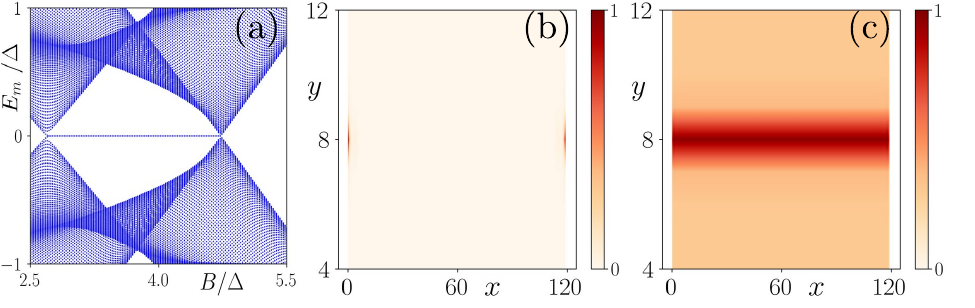}}
	\caption{(a) We illustrate the energy eigenvalue sprctra $E_m$ of the quasi 1D Hamiltonian
	 as a function of $B$ obeying open boundary condition (OBC). In panels (b) and (c), we depict the 
	site-resolved LDOS, computed at $E_m=0$ for the topological ($B/\Delta=3.5$) and the non-topological regime ($B/\Delta=2.0$), respectively. Here, we consider a $120 \times 17$ lattice system. 
	The rest of the model parameters take the values: ($\mu/\Delta, t_{h}/\Delta, \theta$)=(4, 1, 2$\pi/3$).}
	\label{fig:sup_quasi_1D}
\end{figure}
%~~~~~~~~~~~~~~~~~~~~~~~~~~~~~~~~~~~~~~~~~~~~~~~~~~~~~~~~
%------------------------------
\subsection{Dynamical case}
%------------------------------
%~~~~~~~~~~~~~~~~~~~~~~~~~~~~~~~~~~~~~~~~~~~~~~~~~~~~~~~
\begin{figure}[h]
	\centering
	\subfigure{\includegraphics[width=0.58\textwidth]{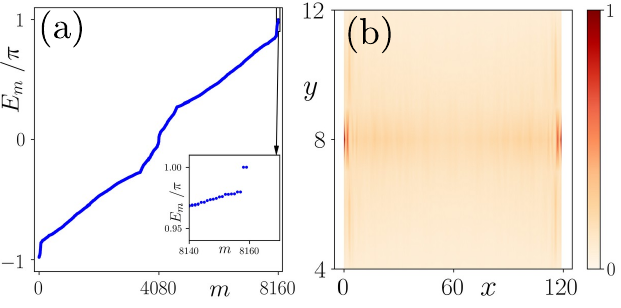}}
	\caption{(a) We depict the quasi-energy eigenvalue spectrum as a function of state index $m$ 
		for the quasi 1D case employing OBC. 
		Two $\pi$-MEMs modes appear in the quasi-energy spectrum and are highlighted in the inset for better clarity. (b) We show site-resolved LDOS for the $\pi$-MEMs to indicate their localization at the two ends of the chain. We choose the model parameters as: ($\mu/\Delta, t_{h}/\Delta, \theta, B/\Delta$, $\Omega/\Delta$)=(4, 1, $2\pi/3$, 2, 3.5).}
	\label{fig:sup_quasi_1D_dyn}
\end{figure}
%~~~~~~~~~~~~~~~~~~~~~~~~~~~~~~~~~~~~~~~~~~~~~~~~~~~~~~
To investigate the Floquet MEMs in a quasi-1D setup ($120 \times 17$ lattice sites), we employ the same driving protocol as discussed in the main text [see Eq.~(2) in the main text]. While starting from the non-topological regime of the above-mentioned static case ($B/\Delta=2$), we obtain anomalous $\pi$-MEMs in the quasi-energy spectrum [see Fig.~\ref{fig:sup_quasi_1D_dyn}(a)]. We also compute the site-resolved LDOS for the $\pi$-modes and illustrate the same in Fig.~\ref{fig:sup_quasi_1D_dyn}(b). One can identify the localized $\pi$-MEMs at the end of the chain from Fig.~\ref{fig:sup_quasi_1D_dyn}(b). Thus, one can generate the dynamical modes employing a quasi-1D setup instead of a strictly 1D system and the essential results remain qualitatively similar. However, to avoid numerical complications 
due to the finite size effect of the quasi-1D system, we resort to a strictly 1D system for all our numerical calculations in the main text.

%~~~~~~~~~~~~~~~~~~~~~~~~~~~~~~~~~~~~~~~~~~~~~~~~~~~~~~~
\begin{figure}[]
	\centering
	\subfigure{\includegraphics[width=0.58\textwidth]{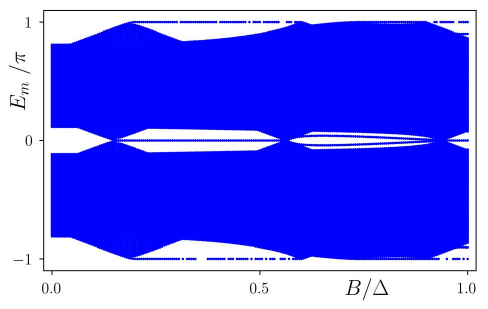}}
	\caption{(a) We illustrate the quasi-energy eigenvalue spectra $E_m$ as a function of the mganetic impurity strength $B$. We choose the model parameters as: ($\mu/\Delta, t_{h}/\Delta, \theta$, $\Omega/\Delta,V_{0}/\Delta$)=(4, 1, $2\pi/3$, 1.5, 5).}
	\label{fig:quasiB}
\end{figure}
%~~~~~~~~~~~~~~~~~~~~~~~~~~~~~~~~~~~~~~~~~~~~~~~~~~~~~~

%\textcolor{red}{
%~~~~~~~~~~~~~~~~~~~~~~~~~~~~~~~~~~~~~~~~~~~~~~~~~~~~~~~~
\section{Discussions} \label{Sec:Exact}
%~~~~~~~~~~~~~~~~~~~~~~~~~~~~~~~~~~~~~~~~~~~~~~~~~~~~~~~~
For the static case, Shiba bands emerge within the parent superconducting gap when magnetic impurity energy scale locally overcomes the superconducting energy scale \ie $B/\Delta \geq 1$. 
The bulk gap within such emegent bands is called the minigap, which exhibits an effective $p$-wave nature~\cite{Felix_analytics}. The closing and re-openning of this minigap give rise to the topological phase transition~\cite{Yazdani2013,Felix_analytics}. In Fig.~\ref{fig:sup_static_winding_number} of SM, for a particular set of the other parameters, we emphasize that the topological phase transition occurs at $B/\Delta \approx 3.18$, giving rise to the emergence of $0$-energy MEMs at the two ends of the magnetic impurity chain.

%\textcolor{red}{%With the same set of the above mentioned parameters, 
In Fig.~\ref{fig:quasiB}, we depict the quasi-energy spectra as a function of the magnetic impurity strength in presence of the external periodic drive. %in Fig.~\ref{fig:quasiB}. 
The emergence of regular $0$- and anamalous $\pi$-modes for $B/\Delta \ll 1$ is clearly visible. The naive explanation behind the early apprearance of $0$-Floquet MEMs (compared to the static case) 
can be attributed to the fact that the application of external drive reduces the bare superconducting gap $\Delta$ by a rough factor of $J_{0}\left(\frac{2 V_{0}}{\Omega}\right)$ (from FPT) resulting in the topological condition $B/\Delta^{{\rm{eff}}}_{{\rm{FPT}}} \geq 1$. However, this condition does not exactly match with the exact numerical results as shown in Fig.~\ref{fig:quasiB}. Moreover, no such explanation can be attributed behind the appearance the anamalous $\pi$-MEMs.

Importantly, the static Shiba minigap $\Delta_{\rm SG}$  scales with the 
%exponentially suppresses the 
parent superconducting gap $\Delta$ in the following way $\Delta_{\rm SG}\propto \Delta \exp(-a/\xi)$ when the coherence length $\xi$ of the parent superconductor is small and $a$ denotes the impurity spacing~\cite{Felix_analytics}. One can naively say that for our case with periodically driven chemical potential, the static Shiba minigap $\Delta_{\rm SG}$ is further modulated by $J_{0}\left( \frac{2 V_{0}}{\Omega}\right)$ $\left[\left( 1- \frac{V_{0}^2}{2 \Omega^2}\right)\right]$ factor, as predicted by FPT [BW] theory. This leads to the double dressed superconducting minigap $\Delta^{\rm F}_{\rm SG}$ 
where $\Delta^{\rm F}_{\rm SG}\approx \Delta_{\rm SG}J_{0}\left( \frac{2 V_{0}}{\Omega}\right) \ll \Delta$ if we follow the FPT, for example. However, this gives us a crude understanding of the
$0$-minigap for the driven case. Moreover, no such explanation can be attributed to the $\pi$-gap hosting anamalous $\pi$-MEMs.

For any perturbative analysis, one can follow two approaches: At first, one can perform perturbative calculation to obtain an effective Floquet Hamiltonain in the presense of both magnetic impurities and external drive, and then analyse the formation of emergent Shiba bands to procure Floquet MEMs. On the other hand, one can also consider Shiba bands for the static system~\cite{Felix_analytics}, and then apply periodic drive in it to realize Floquet MEMs by perturbative analysis. We follow the former approach as this procedure is realistic in which we treat both the periodic drive and the magnetic impurities on equal footing. However, this approach is based on numerical analysis and exact qualitative match with the perturbative analysis (along the same approach) is not apparent as we emphasize in the main text. 
%for practical purpose as the first procedure is realitic as it gives the equal footing to drive and Shiba %band. But we are not getting exactly quatititative match with numerics. 
Probably, to achieve nearly exact qualititative match between the numerical and perturbative analysis, one may has to follow the latter procedure as done for the static case~\cite{Felix_analytics}. 
However, development of such analytically tractable scattering theory in presence of external drive is a challenging task and beyond the scope of the present manuscript. Morover, we believe that any such perturbation theory is inadequate to analytically explain the emergence of Floquet $\pi$-MEMs.
%same as done for static case by F.V. Oppen et al~\cite{Felix_analytics}. However, that is beyond the scope of the current manuscript.
%}	

\end{onecolumngrid}
\end{document}